\documentclass[a4paper,11pt]{article}
\usepackage[utf8]{inputenc}

\title{Classification of molecular sequence data using Bayesian phylogenetic mixture models}

\author{Elisa Loza-Reyes\thanks{Department of Computational and Systems Biology, Rothamsted Research, Harpenden, AL5 2JQ, UK. Email: elisa.loza@rothamsted.ac.uk}, Merrilee Hurn\thanks{Department of Mathematical Sciences, University of Bath, Bath, BA2 7AY, UK.} and Tony Robinson\thanks{Department of Mathematical Sciences, University of Bath, Bath, BA2 7AY, UK.}}

\usepackage{graphicx}   % To use the \includegraphics command
\usepackage{amssymb}
\usepackage{amsthm}
\usepackage{amsmath}
\usepackage{multirow}
\usepackage{subfig}
\usepackage{graphics,color}  % To draw squares alongside text
\usepackage[table]{xcolor}   % To color every alternate row on a table

\usepackage{lineno}

\begin{document}

\maketitle

% Rule to draw squares alongside text
\newsavebox{\myblacksquare}
\savebox{\myblacksquare}{\textcolor{black}{\rule{0.08in}{0.08in}}}

\definecolor{light-gray}{gray}{0.75}
\newsavebox{\mylightgraysquare}
\savebox{\mylightgraysquare}{\textcolor{light-gray}{\rule{0.08in}{0.08in}}}

\definecolor{dark-gray}{gray}{0.40}
\newsavebox{\mydarkgraysquare}
\savebox{\mydarkgraysquare}{\textcolor{dark-gray}{\rule{0.08in}{0.08in}}}

\definecolor{light-light-gray}{gray}{0.97}

\begin{abstract}
\noindent
Rate variation among the sites of a molecular sequence is commonly found in applications of phylogenetic inference. Several approaches exist to account for this feature but they do not usually enable the investigator to pinpoint the sites that evolve under one or another rate of evolution in a straightforward manner. The focus is on Bayesian phylogenetic mixture models, augmented with allocation variables, as tools for site classification and quantification of classification uncertainty. The method does not rely on prior knowledge of site membership to classes or even the number of classes. Furthermore, it does not require correlated sites to be next to one another in the sequence alignment, unlike some phylogenetic hidden Markov or change-point models. In the approach presented, model selection on the number and type of mixture components is conducted ahead of both model estimation and site classification; the steppingstone sampler (SS) is used to select amongst competing mixture models. Example 
applications of simulated data and mitochondrial DNA of primates illustrate site classification via \textquoteleft augmented\textquoteright\, Bayesian phylogenetic mixtures. In both examples, all mixtures outperform commonly-used models of among-site rate variation and models that do not account for rate heterogeneity. The examples further demonstrate how site classification is readily available from the analysis output. The method is directly relevant to the choice of partitions in Bayesian phylogenetics, and its application may lead to the discovery of structure not otherwise recognised in a molecular sequence alignment. Computational aspects of Bayesian phylogenetic model estimation are discussed, including the use of simple Markov chain Monte Carlo (MCMC) moves that mix efficiently without tempering the chains. The contribution to the field of Bayesian phylogenetics is in (1) the use of mixture models augmented with allocation variables as tools for site classification and quantification of 
classification uncertainty, (2) the successful application of SS for selection of phylogenetic mixtures, and (3) the development of novel MCMC aspects of relevance to Bayesian phylogenetic models - whether mixtures or not\footnote{The MCMC methods discussed in this paper have been coded in a C program; source files are available upon request. Supplementary material is available online.}.
\textbf{Key words:} among-site rate variation; Bayesian mixture model; classification; Markov chain Monte Carlo; model selection; phylogeny.
\end{abstract}

\section{Introduction}
\label{sec:intro}

Molecular phylogenetics is the inference and interpretation of evolutionary relations between taxa based on the taxa's DNA or protein sequences. The sequences are arranged on top of one other to form an alignment with as many rows as sequences observed, and roughly as many columns (or \emph{sites}) as characters in the sequences. The conventional likelihood-based model for phylogeny inference (e.g. Felsenstein, 1981\nocite{Felsenstein1981}) contains three parameters of inferential interest: a tree graph which represents the evolutionary relations between the taxa; the branch lengths of this tree which measure the expected number of nucleotide substitutions per site; and a stochastic process which models the evolution of the sequences along the branches of the tree (the latter is usually referred to as the \emph{evolutionary model}). Such a model is complex but may still be too simple to capture important features of the generating process. In particular, it is not uncommon for sites under different 
functional constraints to accumulate substitutions at different rates. It is now well understood that if rate variation among sites is present and is not accounted for by the model, then spurious parameter estimates can be produced (Huelsenbeck and Suchard, 2007 and references therein\nocite{Huelsenbeck_Suchard2007}).

Various approaches have been proposed to account for among-site rate variation in phylogenetic inference, including the gamma model (Yang, 1993; 1994\nocite{Yang1993,Yang1994a}) and several more recent models involving finite mixtures of distributions (e.g. Pagel and Meade, 2004; Lartillot and Philippe, 2004; Huelsenbeck and Suchard, 2007; Webb, Hancock and Holmes, 2009; Evans and Sullivan, 2012\nocite{Pagel_Meade2004,Lartillot_Philippe2004,Huelsenbeck_Suchard2007,Webbetal_2009,Evans_Sullivan2012}). The latter type of models assume that a site is generated from a mixture of multiple processes, each of which may be indexed by a specific tree topology, a specific set of branch lengths and specific parameters of the stochastic evolutionary model. 

Rate variation among sites may be related to quantitative differences in the rates of substitution (e.g. sites with high rates versus sites with low rates) but also to qualitative differences in the pattern of substitution (e.g. sites with large transition/transversion rate ratios versus sites for which all substitution types occur at the same rate; Pagel and Meade, 2004\nocite{Pagel_Meade2004}). In phylogenetic applications it is possible to find quantitative among-site rate variation, qualitative variation, both or neither. 

Developments in phylogenetic mixture modelling have accounted for both types of rate variation and examples of this include Felsenstein and Churchill's approach (1996)\nocite{Felsenstein_Churchill1996}. They account for quantitative variation in substitution rates among sites by a hidden Markov process that operates along the alignment, assigning rates to sites from a finite pool of values. This method incorporates the biologically realistic assumption of correlation between the rates of evolution at consecutive sites, so that the chance of neighbouring sites evolving under the same rate is higher than that of distant sites. A disadvantage of this assumption, however, is that possible biases may be introduced by the removal of sites involving gaps in the alignment, or by other errors that result in consecutive observable sites not being direct neighbours in reality. 

To model qualitative rate heterogeneity, Pagel and Meade (2004)\nocite{Pagel_Meade2004} use a Bayesian mixture of multiple stochastic evolutionary processes. Their model supposes that data at a given site arise from a mixture of multiple classes, each class indexed by a common-to-all-class tree and branch lengths, and a class-specific evolutionary model. Pagel and Meade's mixture assumes a common parametrisation of evolutionary models across components. The assumption of a common set of branch lengths across mixture components results in a phylogeny whose branches are a compromise over the possibly quite different substitutional tempos in the alignment; fast and slow substitution processes are forced into a common medium. This may miss important substitutional heterogeneity and so Pagel and Meade (2008) and Meade and Pagel (2008)\nocite{Pagel_Meade2008, Meade_Pagel2008} consider extensions to their original model, this time allowing for multiple sets of branch lengths. Kolaczkowski and Thornton (2008)\nocite{Kolaczkowski_Thornton2008} present a mixture similar to that in Meade and Pagel (2008), but conduct inference within a maximum-likelihood framework. 

Evans and Sullivan (2012)\nocite{Evans_Sullivan2012} model both quantitative and qualitative rate heterogeneity using mixtures whose components share a common tree topology and set of branch lengths but are indexed by individual evolutionary models, each accompanied by a scaling factor that permits each mixture component to follow its own tempo of substitution. They further allow for different substitution rate constraint cases of the evolutionary GTR model across components and conduct inference on both GTR constraint cases and the number of mixture components via reversible jump MCMC. 

A related approach, called the CAT model (Lartillot and Philippe, 2004\nocite{Lartillot_Philippe2004}), considers qualitative mixtures of stochastic evolutionary processes, all with the same set of substitution rates but with different stationary probabilities. Inference on the number of mixture components is conducted using a Dirichlet process prior and the model is estimated via MCMC. Huelsenbeck and Suchard (2007)\nocite{Huelsenbeck_Suchard2007} consider quantitative mixtures of branch lengths in which sites are partitioned into classes according to a Dirichlet process prior. Sites that are assigned to the same class share a common set of branch lengths, while all sites, irrespective of their class, share a common topology and evolutionary model. Both the number of classes and the assignment of sites to classes are treated as random variables and, together with the usual phylogenetic parameters, are objects of inferential interest.

One aspect of mixture models that has been under-explored in the phylogenetics literature is their use for site classification through the introduction of latent allocation variables. The allocation variables identify the underlying class of a site and thus enable the decomposition of the complicated structure of a mixture into simpler structures. In a phylogenetic context, mixture components may have a direct biological interpretation and site classification can lead to insights of structure and heterogeneity in the alignment that are not otherwise easily uncovered. The purpose of this study is, therefore, to extend the functionality of phylogenetic mixtures to include allocation variables and investigate their use for site classification. The mixture in Pagel and Meade (2004)\nocite{Pagel_Meade2004} is a pioneering example of the use of phylogenetic mixtures for site classification. Their model does not incorporate allocation variables and site classification involves \emph{a posteriori} processing of the 
analysis output to determine the component that is most likely to have generated a site (e.g. recent work by Xi et al., 2012\nocite{Xietal_2012} applies Pagel and Meade's classification approach to interrogate angiosperm molecular sequence data and thus to improve phylogenetic resolution in this group of plants). Lartillot and Philippe (2004) and Huelsenbeck and Suchard (2007)\nocite{Lartillot_Philippe2004,Huelsenbeck_Suchard2007} incorporate allocation variables, but straightforward statements about site classification are obscured by their consideration of the number of mixture components as a random variable. Their MCMC samplers move between mixtures of different dimensions rendering the allocation variables useless for straightforward classification (Lartillot and Philippe, 2004\nocite{Lartillot_Philippe2004}, resort to \emph{a posteriori} multivariate analysis techniques to identify stable components across their MCMC run). More recently, Evans and Sullivan (2012)\nocite{Evans_Sullivan2012} do not 
attempt site classification but acknowledge its importance as a means to extract structural information from the observed data and call it \textquotedblleft a very hard problem made even harder by the estimation of the number of mixture components\textquotedblright.

The approach pursued in this study is to model the class allocation of a site jointly with the rest of the phylogenetic mixture parameters. We do not consider the number of mixture components as a random variable and, in contrast with previous published approaches, determine this parameter using model selection techniques ahead of model estimation. In the absence of label switching, the output from the model estimation stage can be directly used for site classification. We employ the steppingstone sampler (SS) of Xie et al. (2011\nocite{Xieetal_2011}) to estimate the marginal likelihood of a model and thus conduct model selection. We demonstrate that SS is able to correctly recover the true model when applied to simulated data, and further apply it to analyse mitochondrial DNA data. We perform site classification using the model selected by SS, and demonstrate that our use of phylogenetic mixtures augmented by allocation variables correctly detects heterogeneity in the data and accurately classifies the sites to evolutionary components. 
This approach estimates the posterior probability distribution of site allocation so that classification uncertainty may also be appreciated. As part of our MCMC implementation, we present a novel set of move types to update the parameters of a mixture phylogenetic model, and investigate their performance. We show that our MCMC algorithm achieves the same, or greater, efficiency than existing methods with potential for deployment in the estimation of both mixture or non-mixture phylogenetic models at reduced computational cost.

\section{Bayesian phylogenetic mixtures for site classification}
\label{sec:Models}

\subsection{The models}

The backbone of likelihood-based phylogenetic methods is a \emph{homogeneous model} positing that the characters at a site in a DNA alignment are an independent realisation of a continuous-time Markov process, with state space $\mathcal{I}=\{A,C,G,T\}$, that evolves on the branches of a bifurcating tree topology, $\phi$, and has realisations at the leaves of this tree. The instantaneous rate matrix, $Q$, that generates the Markov process is indexed by a (possibly vector) parameter $\theta$. There are several proposed parametrisations of the $Q$-matrix in the literature (e.g. Jukes and Cantor, 1969; Hasegawa et al., 1985\nocite{Jukes_Cantor1969,Hasegawa_etal1985}) with the most general time-reversible one called the GTR matrix (Lanave et al., 1984; Tavar\'{e}, 1986\nocite{Tavare1986,Lanave_etal1984}), where

\begin{equation}
Q(\theta)= \left (
\begin{array}{c c c c}
q_{AA} & r_{AC}\pi_{C} & r_{AG}\pi_{G} & r_{AT}\pi_{T} \\
r_{AC}\pi_{A} & q_{CC} & r_{CG}\pi_{G} & r_{CT}\pi_{T} \\
r_{AG}\pi_{A} & r_{CG}\pi_{C} & q_{GG} & r_{GT}\pi_{T} \\
r_{AT}\pi_{A} & r_{CT}\pi_{C} & r_{GT}\pi_{G} & q_{TT}
\end{array} \right )
\end{equation}

\noindent and $\theta=(r,\pi)$ is a collection of six substitution rates $r=(r_{AC},\ldots,r_{GT})$ and four stationary probabilities $\pi=(\pi_A,\ldots,\pi_T)$ with constraints $r_m,\pi_i\geq 0$ and $\sum r_m = \sum \pi_i =1$ ($m=AC,\,AG,\,\ldots,\,CT,\,,GT$; $i=A,\,C,\,G,\,T$). The diagonal values of $Q$ are defined so that each row adds up to zero. The expected total rate of substitution of the process generated by matrix $Q$ is equalled to one (Felsenstein, 1981\nocite{Felsenstein1981}) so that the branch lengths represent the expected number of substitutions per site. The Markov process of character substitution is time-reversible, a feature that prevents us from inferring rooted trees. Thus, for an observed alignment of size $S$ sequences $\times$ $N$ sites, parameter $\phi$ takes values in the set of unrooted bifurcating leaf-labelled trees for $S$ taxa; branch lengths are real valued; and the space in which parameter $\theta$ takes values is dictated by the chosen parametrisation of matrix $Q$. The objective of the analysis is usually inference about the tree topology, $\phi$, this tree's branch lengths (denoted by a set $t=\{t_{1},\ldots,t_{2S-3}\}$) and $\theta$.

Building upon the homogeneous model, we account for among-site rate variation using a finite mixture of distributions of the type

\begin{equation}
x_n \mid \omega , \phi , t , \theta , k \sim\sum_{j=1}^{k}\omega_{j}\,p(x_{n}\mid \phi,t_j,\theta_j),\hspace{0.25cm}\mbox{independently for }n=1,\ldots,N,
\label{eq:mixture}
\end{equation}

\noindent where $x_n$ is the observed data at site $n$; $k$ is the number of mixture components; $\omega = ( \omega_{1},\ldots,\omega_{k} )$ are the mixture proportions ($\omega_j\geq 0$ and $\sum_{j=1}^{k}\omega_j=1$); each component $j$ $(j=1,\ldots,k)$ has set of branch lengths $t_j$ and parameters of the $Q$-matrix $\theta_j$ collectively denoted by $t=(t_1,\ldots,t_k)$ and $\theta = ( \theta_1,\ldots,\theta_k)$; and $p(x_{n}\mid \phi,t_1,\theta_1),\ldots,p(x_{n}\mid \phi,t_k,\theta_k)$ are the $k$ component likelihoods. Model~\eqref{eq:mixture} thus asserts that characters at site $n$ are generated from a mixture of $k$ different evolutionary components occurring in proportions $\omega_1,\ldots,\omega_k$. To decompose the structure of this mixture, we extend the model to include a set of $N$ scalar latent allocation variables, $z = (z_1,\ldots,z_N)$, where each $z_n \in \{1,\ldots,k\}$. Extending Bayesian mixture models to include allocation variables is fairly standard in the statistics literature (e.g.
 Richardson and Green, 1997\nocite{Richardson_Green1997}), sometimes for their own sake when classification is the goal and other times just because they can simplify the computational fitting. The $n^{th}$ of these latent variables answers the question ``to which component does observation $n$ belong?''.
Including $z$ in the model, we can write

\begin{equation}
x_n\mid \phi, t, \theta, k, z_n\sim p(x_{n}\mid \phi,t_{z_n},\theta_{z_n}),\hspace{0.25cm}\mbox{independently for }n=1,\ldots,N.
\label{eq:mixture2}
\end{equation}

This formulation not only accounts for both quantitative and qualitative rate heterogeneity, but also provides a means to class discovery by the use of $z$. In addition to classifying the sites to evolutionary components, mixture~\eqref{eq:mixture2} also enables us to discern the profile of each class by estimating the component-specific parameters. So, the analysis may lead to statements such as \textquotedblleft class $1$ is more conserved than class $2$ as the former displays a shorter total branch length than the latter\textquotedblright\, or \textquotedblleft the nucleotide composition of the two classes is quite different, as reflected by the estimated stationary probabilities\textquotedblright.

The joint prior of all parameters is expressed as

\begin{eqnarray}
p(\omega,z,\phi,\theta,t) &=& p(\omega)p(z\mid\omega)p(\phi \mid z,\omega)p(\theta \mid \phi,z,\omega)p(t \mid \theta,\phi,z,\omega) \nonumber 
\nonumber \\
		 &=& p(\omega)p(z\mid \omega)p(\phi)p(t)p(\theta)
\label{eq:full_model}
\end{eqnarray}
\noindent

\noindent where we have suppressed the explicit conditioning on $k$ because we consider only mixtures with a fixed number of components and make independence assumptions between all parameters other than $z$ and $\omega$. The prior for $\omega$ is taken to be the symmetric Dirichlet distribution $\omega \sim Dir_k(\rho,\ldots,\rho)$. We express prior ignorance about class size by setting $\rho=1$. 

Conditional on $\omega$, the allocations $z_1,\ldots,z_N$ are assumed independent and identically distributed 

\begin{equation}
Pr(z_{n}=j\mid \omega)=\omega_{j},\hspace{0.9cm}j=1,\ldots,k.
\label{eq:prior_allocs}
\end{equation}

We make the following standard choices for the priors on phylogenetic parameters. All tree topologies are assumed to be equally likely \emph{a priori}; that is, we take a discrete uniform prior for $\phi$ (e.g. Suchard et al., 2001\nocite{Suchard_etal2001}). The prior distribution for branch lengths makes an assumption that the $2S-3$ branches for each of the $k$ components are independent both within components and across components. Exponential priors on individual branch lengths are specified, with exponential-rate parameter $\eta$ so that $\mathtt{E}(t_{h,j})=1/\eta$ for branch length $h$ in the $j$th mixture component ($h=1,\ldots,2S-3$; $j=1,\ldots,k$). For the parameter vectors $\theta$ of the $k$ instantaneous rate matrices, we assume independent prior distributions on each $r_j$ and $\pi_j$ of the form $r_j \sim Dir_6(1,\ldots,1)$ and $\pi_j \sim Dir_4(1,\ldots,1)$.

Throughout, the model specified by~\eqref{eq:mixture} and~\eqref{eq:mixture2} is referred to as the \emph{$Q+t$ mixture model}.

We also consider nested submodels of the $Q+t$ mixture. Firstly, we consider mixtures of multiple $Q$ matrices which share a common set of branch lengths, $t$, and tree topology, $\phi$, (Pagel and Meade, 2004\nocite{Pagel_Meade2004}):

\begin{equation}
x_n \mid \omega , \phi , t , \theta , k \sim\sum_{j=1}^{k}\omega_{j}\,p(x_{n}\mid \phi,t,\theta_j),\hspace{0.25cm}\mbox{independently for }n=1,\ldots,N.
\label{eq:mixture_Q}
\end{equation}

Restricting this further we consider mixtures of branch lengths and $Q$ matrices, but where the $Q$ matrices across components share the same stationary probabilities, i.e. $\theta_1=(r_1,\pi),\ldots,\theta_k=(r_k,\pi)$:

\begin{equation}
x_n \mid \omega, \phi, t, \theta, k \sim\sum_{j=1}^{k}\omega_{j}\,p(x_{n}\mid \phi,t_j,r_j,\pi),\hspace{0.25cm}\mbox{independently for }n=1,\ldots,N.
\label{eq:mixture_r+t}
\end{equation}

Both models can be augmented with allocation variables. We refer to model~\eqref{eq:mixture_Q} and its corresponding augmented formulation as the \emph{$Q$ mixture}, and to model~\eqref{eq:mixture_r+t} and its augmented version as the \emph{$r+t$ mixture}. For simplicity, we use the notation $Q+t(k)$ to denote a $Q+t$ mixture with $k$ components, and similarly for $Q(k)$ and $r+t(k)$.

Our mixture models share certain similarities with those used in Pagel and Meade (2004, 2008), Meade and Pagel (2008), Lartillot and Philippe (2004), Huelsenbeck and Suchard (2007) and Evans and Sullivan (2012) but differ in one key way. They are formulated to include allocation variables and the posterior probability of these variables is an object of primary inferential interest and a means for both site classification and quantification of classification uncertainty.

\subsection{Likelihood computation}
\label{sec:Likelihood}

The likelihood function under the most general $Q+t$ mixture is the product of the distributions at individual sites (equation \eqref{eq:mixture2}), from site $1$ to $N$:

\begin{equation}
L(\phi,t,\theta | x,z) = \prod_{n=1}^{N}p(x_n |  \phi,t_{z_n},\theta_{z_n}).
\label{eq:likelihood_Q+t}
\end{equation}

We assume that substitutions at different branches of the tree and among different sites in the alignment are independent of one another. Likelihood~\eqref{eq:likelihood_Q+t} is usually computed for a specific tree and so each tree topology requires a reformulation of this function according to its corresponding branching structure; the larger the tree the more computationally prohibitive the calculation. A recursive technique for the efficient computation of phylogenetic likelihood functions, called the pruning algorithm, was introduced by Felsenstein (1981),\nocite{Felsenstein1981} and this is the algorithm that we use.

\section{Model estimation}
\label{sec:MCMC_sampler}

Markov chain Monte Carlo (MCMC) will be required to fit models of this complexity and we present the basic move types in our MCMC sampler. A distinctive feature of our method is that changes to the topology are separated from those in branch lengths; this is particularly important for some of the mixtures where the components share a common topology but have different sets of branch lengths. Metropolis-Hastings methods are equally valid when the available moves are scanned either randomly or systematically. Here, we have chosen to take the latter approach making use of six move types:

\begin{flushleft}
(a) updating the tree topology $\phi$;\\
(b) updating all branch lengths $t_1,\,t_2,\,\ldots,\,t_{2S-3}$;\\
(c) updating the vector of substitution rates $r$;\\
(d) updating the vector of stationary probabilities $\pi$;\\ 
(e) updating the vector of mixture proportions $\omega$;\\
(f) updating all site allocations $z_1,\,z_2,\,\ldots,\,z_N$.
\end{flushleft}

\noindent Moves (b) -- (d) are applied to all components in the mixture, if required. One complete pass over these six moves is an iteration, the basic time step of our MCMC sampler. The first two move types focus on the tree while the next two concentrate on the parameters of the models on the tree; the last two move types concern the mixture allocations and proportions. We now consider the three groups separately in the context of the most general $Q+t$ mixture model.

\subsection{Updating the tree topology and branch lengths}

The tree topology is updated via the nearest neighbour interchange (NNI) (Robinson, 1971;\nocite{Robinson1971} Moore, Goodman and Barnabas, 1973\nocite{Moore_etal1973}), in which one of the two nearest neighbours of the current topology (in NNI space) is proposed with equal probability. NNI generates a candidate topology while preserving the current set of branch lengths. A candidate topology $\phi^\prime$ is accepted with a probability that simplifies to:

\begin{equation}
a(\phi,\phi^\prime) = \min \left \{1,\;\frac{L(\phi^\prime,t,\theta \mid  x,z)}{L(\phi,t,\theta \mid  x,z)} \right \}.
\label{eqn:MHacceptance_topology}
\end{equation}

A separate proposal mechanism is used to update the branch lengths while maintaining the same topology. We consider two different proposals for branch lengths:

\begin{itemize}
\item Branch length multiplier (BLM).
Also known as proportional shrinking and expanding (Yang, 2006\nocite{Yang2006}), this proposal updates the length of a randomly chosen branch $t_{h,j}$ by multiplying it by a quantity $m$ generated from the density
\begin{equation}
f(m) = (\lambda\,m)^{-1},\hspace{1.5cm}1/\delta<m<\delta
\end{equation}

\noindent where $\lambda=2\mbox{ log }\delta$ and $\delta>1$ acts as a tuning parameter.

\item Branch length normal additive (BLNA).
Also known as the sliding window proposal (Huelsenbeck and Ronquist, 2001\nocite{Huelsenbeck_Ronquist2001}), this mechanism updates a randomly chosen branch length $t_{h,j}$ via an additive Gaussian perturbation, $t_{h,j}^\prime \sim N(t_{h,j},\sigma^2)$, so that $\sigma^2$ acts as the tuning parameter. If negative branch lengths are proposed, they are reflected at zero with the proposal still remaining symmetric.
\end{itemize}

BLM may be thought of as self-tuning as the variance of the proposed branch length is proportional to the square of the original length. This works well when exploring large branches but can be a bit sticky when branch lengths are small as it can take a large number of iterations to move a short distance. On the other hand, a candidate branch length generated from the BLNA proposal has a step size which depends only on the tuning parameter $\sigma^2$ and not on the current branch length. This makes it hard for BLNA to work equally effectively at both large and small scales. In experiments, we achieved best performance by alternating between BLM and a BLNA tuned for small branch lengths (Supplementary Material I).

The acceptance probability of a branch length proposed from either BLM or BLNA is

\begin{equation}
a( t_{h,j},t_{h,j}^\prime )  = 
\min \left \{1,\;\frac{p(t_{h,j}^\prime)}{p(t_{h,j})}\;\frac{L(\phi,t^\prime,\theta \mid  x,z)}{L(\phi,t,\theta \mid  x,z)}\;\frac{q(t_{h,j}^\prime,t_{h,j})}{q(t_{h,j},t_{h,j}^\prime)}\right \}.
\label{eq:MHacceptance_branchBLNA}
\end{equation}

\noindent The proposal ratio $q(t_{h,j}^\prime,t_{h,j})/q(t_{h,j},t_{h,j}^\prime)$\, simplifies to $m$\, for BLM and to $1$ for BLNA, and so acceptance $a( t_{h,j},t_{h,j}^\prime )$\, simplifies to $m\,e^{-\eta(t^\prime_{h,j}-t_{h,j})}$\, and to $e^{-\eta(t^\prime_{h,j}-t_{h,j})}$ for BLM and BLNA, respectively, times the likelihood ratio in both cases.

\subsection{Updating the Markov process parameters}

The $j$th component of the $Q+t$ mixture has a set of parameters controlling the substitution rates plus a set of stationary probabilities, $r_{AC,j},\ldots,r_{GT,j}$ and $\pi_{A,j},\ldots,\pi_{T,j}$, respectively. Since we can treat each mixture component separately for updating purposes, we drop the subscript $j$. Both types of parameters are constrained to sum to one and, as they utilise the same type of proposal, here we concentrate on the substitution rates.

We generate a new set of substitution rates, $r^\prime$, from a Dirichlet distribution centred at the current rate values with a positive shift $\epsilon>0$ and with tuning parameter $\alpha >0$; i.e. $r^\prime \sim Dir_6(\alpha\,(r_{AC}+\epsilon),\ldots,\alpha\,(r_{GT}+\epsilon))$. The variance of the $m$th element of a rate vector proposed with this move type (henceforth referred to as the $\epsilon$Dirichlet proposal) is:

\begin{equation}
var(r^\prime_m) =
\frac{\alpha(r_m+\epsilon)(\alpha_0 - \alpha(r_m+\epsilon))}{\alpha_0^2(\alpha_0+1)}
\label{eq:exp_Dirichlet_corrected}
\end{equation}

\noindent where $\alpha_0=\sum_{m=AC}^{GT} \alpha(r_{m}+\epsilon)=\alpha\,(1+6\epsilon)$. When $\epsilon=0$, our proposal becomes the popular Dirichlet proposal by Larget and Simon (1999\nocite{Larget_Simon1999}), which suffers from one major drawback: when $r_m$ is close to zero so too is $var(r^\prime_m)$. This can create an undesirable cycle in which the MCMC sampler keeps proposing candidate rates very close to zero because the step size of the proposal is nearly zero, typically needing many iterations to escape. We introduce the offset $\epsilon$ as an effective way to improve the mixing of the chain without resorting to tempered schemes that can result in high computational burden. The effectiveness of this proposal is investigated in Supplementary Material II. The move is accepted with probability

\begin{equation}
a( r,r^\prime) = \min \left \{1,\;\frac{L(\phi,t,\theta^\prime \mid  x,z)}{L(\phi,t,\theta \mid  x,z)}\;\frac{q(r^\prime,r)}{q(r,r^\prime)} \right \}
\label{eqn:MHacceptance_rates_frequencies}
\end{equation}

\noindent where the proposal ratio $q(r^\prime,r)/q(r,r^\prime)$ is calculated as the quotient of two Dirichlet density functions.

\subsection{Updating the mixture parameters}

Updating the allocation variables and the vector of mixture proportions is a fairly standard problem in the estimation of Bayesian mixtures via MCMC. The vector of mixture proportions $\omega=(\omega_{1},\ldots,\omega_{k})$ is usually updated using a Gibbs sampler since their posterior conditional is easily seen to be a Dirichlet distribution with parameters $\rho+N_1,\ldots,\rho+N_k$, where $N_j=\sum_{n=1}^{N}I[z_n=j]$ is the number of sites allocated to component $j$ and $I[\cdot]$ is the indicator function. This mechanism thus updates $\omega$ according to the number of sites allocated to each component on a given MCMC iteration. A well known difficulty of this proposal is that it may mix badly when one or more components become quite small or when the other parameters characterising the components make it hard for a site to swap components (see Leslie, 2007\nocite{Leslie07} or Hurn et al., 2008\nocite{Hurn08} for examples in quite different application areas). In the latter case, Leslie (2007)\nocite{Leslie07} and Hurn et al. (2008)\nocite{Hurn08} both suggest a strategy that updates $\omega$ and the allocations jointly. However here we are primarily worried about instances in the MCMC path when one or more components become quite small, causing the chain to mix badly. Given our experience in updating $r$ and $\pi$, we use a shifted Dirichlet approach, here replacing the Gibbs draw from a $Dir_k(\rho+N_1,\ldots,\rho+N_k)$ by a Metropolis-Hastings proposal, $\omega^\prime \sim Dir_k(\rho+N_1+\epsilon,\ldots,\rho+N_k+\epsilon)$ with $\epsilon>0$. The acceptance probability of this move type simplifies to 

\begin{equation}
a( \omega , \omega^\prime )  = 
\min \left \{1 , \prod_{j=1}^k \left( \omega_j / \omega^\prime_j\right)^{\epsilon} \right \} 
\end{equation}

\noindent and so a high acceptance rate is maintained for small values of $\epsilon$.

The allocation for the $n$th site, $z_n$, is updated by drawing, randomly and with equal probability, from the set $\{1,\ldots,k\}_{-z_{n}}$ where the subindex denotes that $z_{n}$ is excluded from the set. Since allocations are updated one at a time, the acceptance probability involves a ratio of likelihoods only at site $n$:

\begin{equation}
a(z_n,z^\prime_n) = \min \left \{1,\;\frac{\omega_{z^\prime_n}}{\omega_{z_n}}\;\frac{p(x_n \mid \phi,t_{z^\prime_n},\theta_{z^\prime_n})}{p(x_n \mid \phi,t_{z_n},\theta_{z_n})} \right \}.
\end{equation}

\subsection{The label switching problem}
\label{sec:label_switch}

The application of MCMC methods to mixture models suffers from the problem that the components of a mixture are intrinsically non-identifiable since the same likelihood function is obtained for any permutation of the component labels.  If the prior distributions do not contain information regarding labelling of the components, this may produce MCMC output that switches labelling throughout the run and masks component-specific information.  In that case, label switching needs to be dealt with before any meaningful inference can be made on parameters for individual mixture components.  Our classification method relies on component-specific inferences and so, if label switching is detected, it is necessary to deal with it.  A description of this phenomenon, together with a review of existing solutions, can be found in Jasra et al., (2005)\nocite{Jasra_Holmes_Stephens2005}.  We note that the commonly used approach of imposing an ordering on one of the parameters, used for example in Richardson and Green (1997)\nocite{Richardson_Green1997}, can be applied after running the simulations when it is clearer which parameter is most effective to use for unpicking the label switching for inference purposes.

\section{Model selection via steppingstone sampling}

% Model choice using Bayes factors and the marginal likelihood
We now turn to the choice of which model to use for a particular set of data. Bayes factors (BFs) can be computed to summarise the evidence provided by the data in favour of one model relative to another (Kass and Raftery, 1995)\nocite{Kass_Raftery1995}. When two models are equally likely \emph{a priori}, the BF is defined as the ratio of the marginal likelihood under model $M_1$ to the marginal likelihood under a second model, $M_0$, given the data, $x$. BFs are usually interpreted on the log scale using the rule of thumb that $2ln(\mbox{BF})>10$ indicates very strong evidence in favour of model $M_1$, $0\leq 2ln(\mbox{BF})\leq2$ indicates no significant difference between the models, and with a range of levels in between according to a scale provided in Kass and Raftery (1995)\nocite{Kass_Raftery1995}. 

The Savage-Dickey ratio (Verdinelli and Wasserman 1995; Suchard et al. 2001\nocite{Verdinelli_Wasserman1995,Suchard_etal2001}) has been successfully used in phylogenetics to estimate BFs directly. This approach requires that models are nested and, although potentially useful for some models in our problem, we have not used this approach here. There exist a number of other ways to estimate BFs; Friel and Pettitt~(2008)\nocite{friel2008marginal} categorise these broadly into {\it across model} or {\it within model} approaches. The most common examples of the former type use reversible jump MCMC (RJMCMC; Green, 1995\nocite{green1995reversible}), embedding the entire set of possible models within a larger model and then sampling from this joint posterior of both models and parameters. Under this approach, BFs of pairs of the competing models are estimated by the relative proportions of time the sampler spends in the two models. Within model approaches, on the other hand, consider the competing models one by one, 
estimating each marginal likelihood in turn in order to estimate the BFs. Friel and Pettitt~(2008)\nocite{friel2008marginal} suggest that the across model approach is the better suited of the two in situations where there are many competing models. Set against this though is the well documented difficulty in constructing efficient RJMCMC algorithms, combined with the associated long run times required to achieve stable estimates. This is particularly a problem when the competing models do not relate to one other in a way that suggests natural moves between their sets of parameters. 

% Approaches to marginal likelihood estimation
We have opted to follow a within model approach and so need to estimate the marginal likelihood for each competing model. The marginal likelihood for model $M_i$ is the expectation (under the prior) of the likelihood of the data $x$, conditioned on the model $M_i$ (or, equivalently, the integral over the parameters of the joint distribution of the data and the prior conditioned on the model),

\begin{equation}
p(x\mid M_i)=\int_{\vartheta_i} p(x\mid \vartheta_i, M_i)p(\vartheta_i\mid M_i)d\vartheta_i
\label{eqn:marglik}
\end{equation}

\noindent where $\vartheta_i$ is the parameter vector of model $M_i$. The marginal likelihood in equation~\eqref{eqn:marglik} cannot be calculated analytically except for the most elementary applications, and its estimation is the topic of considerable interest (e.g. recent work by Ardia et al. (2012)\nocite{Ardia_etal2012} reviews popular Monte Carlo methods for marginal likelihood estimation). An unfortunately commonly used estimator is the harmonic mean (HM) of Newton and Raftery (1994)\nocite{Newton_Raftery1994} which is a form of importance sampling, taking the posterior as its importance distribution. HM marginal likelihoods can be calculated from the MCMC chain used for fitting the model at little extra cost but generally perform poorly. Two alternatives that have been used in the phylogenetic literature are thermodynamic integration (Lartillot and Philippe, 2006\nocite{Lartillot_Philippe2006}; Friel and Pettitt, 2008\nocite{friel2008marginal}) and steppingstone sampling (SS) in either its original (Xie et al., 2011\nocite{Xieetal_2011}) or generalised (Fan et al., 2011\nocite{Fanetal_2011}) flavour.

Marginal likelihood estimation via the original SS method has been shown to be more accurate than both thermodynamic integration and HM in applications to Bayesian phylogenetics (Xie et al., 2011; Baele et al., 2012\nocite{Xieetal_2011,Baeleetal_2012}), while the generalised flavour of SS improves upon its original version in terms of efficiency and stability. Nevertheless, generalised SS requires the specification of a reference distribution that approximates the posterior of interest; for complex phylogenetic mixture models it is unclear how such a reference can be chosen. 

%Stepping-stone sampling
The original SS method (henceforth referred to simply as SS) estimates the marginal likelihood from several MCMC runs stitched together along a path that goes from the posterior to the prior with differing power posterior distributions in between. The power posterior distribution under model $M_i$ and with parameter vector $\vartheta_i$ is:

\begin{equation}
p_{\beta}(\vartheta_i \mid x)=\frac{p(x \mid \vartheta_i)^{\beta}\,p(\vartheta_i)}{c_{\beta}}
\label{eqn:power_posterior}
\end{equation}

\noindent where $0\le \beta \le 1$ and $c_{\beta}$ is a normalising constant. Dependence on the model under consideration ($M_i$) has been suppressed in the notation for simplicity. When $\beta=1$, the power posterior is the posterior distribution and the normalising constant $c_1$ is the marginal likelihood, i.e. $c_1=p(x)$. The power posterior is equivalent to the prior distribution when $\beta=0$ and, assuming that the prior is proper, $c_0=1$. The basic idea of SS is to express the marginal likelihood as the product of $K$ ratios:

\begin{eqnarray}
p(x) &=& \frac{c_1}{c_0} \nonumber \\ \vspace{1cm}
     &=& \Bigl(\frac{c_{\beta_{1}}}{c_{\beta_{0}}} \Bigr)\,\ldots\,\Bigl(\frac{c_{\beta_{\nu}}}{c_{\beta_{\nu-1}}} \Bigr)\,\ldots\,\Bigl(\frac{c_{\beta_{K}}}{c_{\beta_{K-1}}} \Bigr).
\label{eqn:marginal_likelihood_SS}
\end{eqnarray}

\noindent where $0=\beta_0<\ldots<\beta_{\nu-1}<\beta_{\nu}<\ldots<\beta_K=1$ are the stepping stones between the prior and the posterior distributions. Each ratio $c_{\beta_{\nu}}/c_{\beta_{\nu-1}}$ is estimated as the average value of the observed likelihoods raised to the power $\beta_{\nu}-\beta_{\nu-1}$ when sampled from a MCMC run with target distribution $p_{\beta_{\nu-1}}$. Therefore, SS does not require samples from the posterior. In practice, however, we start by sampling from the posterior to burn-in the chain and proceed in the direction $p_{\beta_{K-1}}$ until reaching the prior $p_{\beta_{0}}$. 

% Steppingstone sampling in practice
To construct the SS sampler from our MCMC algorithm, the proposals remained unchanged and the likelihood ratios in acceptance probabilities $a(\phi,\phi^\prime)$, $a(t_{h,j},t^\prime_{h,j})$, $a(r,r^\prime)$ and $a(z_n,z^\prime_n)$ were raised to the power $\beta_{\nu-1}$. We set $K=30$ and spaced the values of $\beta$ according to uniform quantiles of a $Beta(0.3,\,1)$ distribution which, in practice, entailed setting $\beta_{\nu}=(\nu/K)^{3.33}$. Xie et al. (2011\nocite{Xieetal_2011}) show that the accuracy of SS is optimal in a Gaussian model example when $\beta$ values are set in this way. Assessing optimal specification of $\beta$ in Bayesian phylogenetic mixture applications is outside the scope of this study so we followed the Xie et al. (2011\nocite{Xieetal_2011}) recommendation. However, see also some recent work by Friel, Hurn and Wyse (2013\nocite{friel2012improving}).

We employed SS to select between mixture models with differing parameterisations (e.g. $Q+t$ versus $Q$) and also to assess the number of components in a given mixture (e.g. $k=2$ versus $k=3$). The aim of model selection is not necessarily to find the true model that generated the data but to select a model that captures the key features of the data while being biologically realistic and tractable (Steel, 2005\nocite{Steel_2005}). Only once we have determined the most plausible mixture and its number of component subpopulations for a given data set, do we perform site classification.

\section{Classification of simulated data}
\label{sec:Simulation}

% (1) Introduce synthetic data set
\subsection{Methods}
To validate our classification approach, we generated a synthetic DNA alignment of size $16$ sequences $\times\, 2\,500$ sites, with the software package Seq-Gen (Rambaut and Grassly, 1997~\nocite{Rambaut_Grassly1997}). Sites $1-1500$ were generated from an evolutionary class with substitution rates $\{r_{AC}\,=\,r_{AT}\,=\,r_{CG}\,=\,r_{GT}\,=\,0.0500,\,r_{AG}\,=\,r_{CT}\,=\,0.4000\}$, stationary probabilities $\{\pi_A=0.3220,\,\pi_C=0.3040,\,\pi_G=0.1080,\,\pi_T= 0.2660\}$ and total branch length $T=10$. Sites $1501-2500$ were simulated with $\{r_{AC}=0.1009,\,r_{AG}=0.3645,\,r_{AT}=0.1506,\,r_{CG}=0.0639,\,r_{CT}=0.3044,\,r_{GT}=0.0157\}$; $\{\pi_A=\ldots=\pi_T=0.2500\}$ and $T=0.1$. Both classes were generated under the same tree topology, which was randomly sampled from the space of all unrooted bifurcating trees that relate $16$ sequences. In our experiments, the topology was held fixed at its generating value.

% (2) Establish relevance of the synthetic data set
The intention here is to assess whether the classification method is able to detect the substitutional differences between the two classes and to correctly allocate sites to evolutionary groups without prior knowledge of the partitioning in the data. 

% (3) Describe results 
% (3.1.) Model selection
\subsection{Model selection}

Before the runs for inference, we conducted several exploratory runs to tune the SS proposals to $\delta=1.5$ for the BLM move; $\sigma=0.08$ for BLNA; $\alpha_r=900$; $\alpha_{\pi}=700$ and $\epsilon_{\theta}=0.0001$ for $\epsilon$Dirichlet of substitution rates and stationary probabilities; and $\epsilon_{\omega}=0.0001$ for $\epsilon$Dirichlet of mixture proportions. Hyperparameter $\eta$ for the prior on a branch length was set to $4.5$ as a compromise between the two simulated classes. The SS sampler was run for $5\,000$ iterations for each $\beta$ value in the steppingstone path. The burn-in phase consisted of $20\,000$ iterations at power $\beta=1$ (the posterior). It is worth noting that one iteration in our MCMC sampler systematically updates all the parameters in the model. So, there are $2+k(2S-1)+N$ parameter updates per iteration in a fit of a $Q+t$ mixture with $k$ components to an alignment of $S$ sequences and $N$ sites (one update for the topology, $k(2S-3)$ for all branch lengths across all 
mixture components, $2k$ for the rate and stationary probability vectors across all components, $N$ for all site allocations and one for the vector of mixture proportions). Care must be therefore taken when assessing the length of our runs; $5\,000$ iterations here correspond to $12.8\times10^6$ parameter updates.

\begin{figure} [p] \centering
  \subfloat[]
      {\includegraphics[scale=0.45]{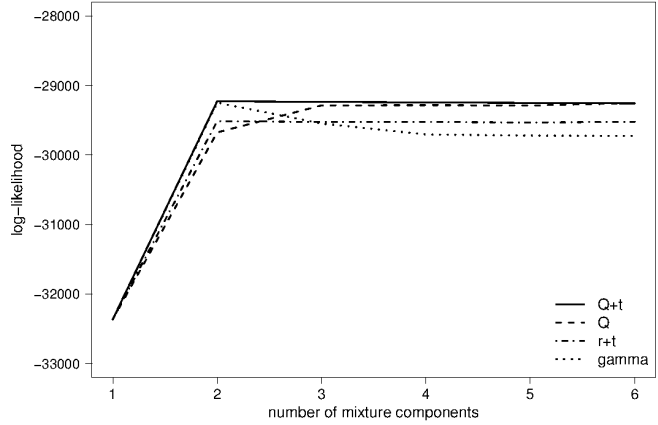}}
  \vspace{0.3cm}
  \subfloat[]
      {\includegraphics[scale=0.45]{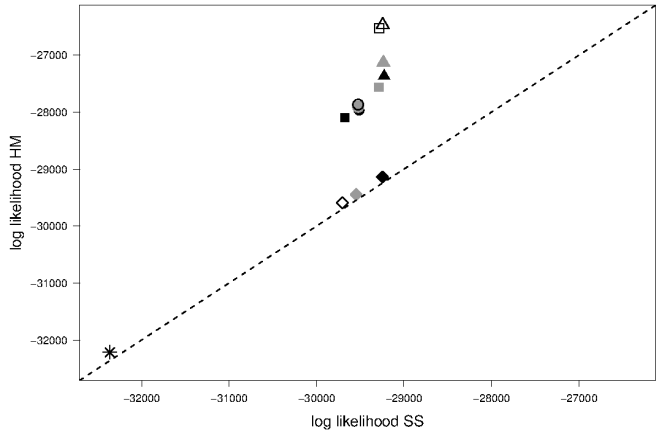}}
\caption{(a) Estimated log marginal likelihoods for models considered for the synthetic DNA alignment. Plotted data: homogeneous: $-32\,364$; $Q+t$ with 2--6 components: $-29\,226$, $-29\,235$, $-29\,243$, $-29\,250$, $-29\,257$; $Q$ with 2-6 components: $-29\,675$, $-29\,287$, $-29\,282$, $-29\,288$, $-29\,253$; $r+t$ with 2--6 components: $-29\,513$, $-29\,524$, $-29\,525$, $-29\,532$, $-29\,522$; gamma with 2--6 categories: $-29\,246$, $-29\,546$, $-29\,704$, $-29\,720$, $-29\,726$. (b) Scatter plot of log marginal likelihood values estimated using HM versus SS. Plotted values on the x-axis are given in legend of Fig. 1(a). Plotted values on the y-axis (geometric symbols in black, grey and white fill correspond to 2, 3 and 4 model components/categories, respectively): homogeneous (*): $-32\,216$; $Q+t$ ($\bigtriangleup$) with 2--4 components: $-27\,369$, $-27\,134$, $-26\,467$; $Q$ ($\Box$) with 2--4 components: $-28\,099$, $-27\,563$, $-26\,530$; $r+t$ ($\bigcirc$) with 2--4 components: $-27\,966$, $-27\,
908$, $-27\,868$; gamma ($\diamondsuit$) with 2--4 categories: $-29\,138$, $-29\,447$, $-29\,594$. Dashed line: region in which HM and SS estimates would agree.}
% Difference between the log-marginal-likelihoods estimated by HM and SS for a range of models. Plotted data: difference homogeneous model: $145$; $Q+t$ mixture with 2--4 components: $1\,857$, $2\,101$, $2\,776$; $Q$ mixture with 2--4 components: $1\,576$, $1\,723$, $2\,752$; $r+t$ mixture with 2--4 components: $1\,547$, $1\,616$, $1\,657$; gamma: $123$.
\label{fig:SYNTHETIC_ModelChoice}
\end{figure}

Models~\eqref{eq:mixture},~\eqref{eq:mixture_Q} and~\eqref{eq:mixture_r+t}, in their allocation-variable formulation, were considered for the synthetic alignment with $k=1,\ldots,6$ components. Figure~\ref{fig:SYNTHETIC_ModelChoice}(a) shows the log marginal likelihoods for these models, estimated using the SS sampler. The log-likelihood for $k=1$ is common across all mixture types and it corresponds to a fit of the data with the homogeneous model. It is clear that any mixture fits these data better than the homogeneous model, which is unsurprising given the heterogeneity that underlies the alignment. For comparison, we also estimated the log marginal likelihood of a special type of mixture model in which all components share one $Q$ and set of branch lengths, but each component is allowed a separate scalar $\gamma_j$ that scales the $Q$ matrix and that is drawn from a discrete version of a gamma distribution with empirically-estimated shape parameter (Yang, 1993; 1994\nocite{Yang1993,Yang1994a}; see also 
Pagel and Meade, 2004\nocite{Pagel_Meade2004} for a description in the context of mixture models). The log marginal likelihood of the discrete-gamma model, as this formulation is usually known, was estimated for $2-6$ categories of the discrete gamma distribution using the software package MrBayes 3.2 (Ronquist et al., 2012\nocite{Ronquist_etal2012}). We specified the number of MCMC cycles and settings so that MrBayes' analysis was comparable to our SS sampler. The discrete-gamma model is a popular choice in phylogenetics because it accounts for rate heterogeneity in an elegant and convenient way by adding only one extra parameter to the homogeneous model formulation; the shape parameter of the gamma distribution. But such convenience and elegance may be insufficient when dealing with complex evolutionary scenarios. Figure~\ref{fig:SYNTHETIC_ModelChoice}(a) shows that the two best performing models are $Q+t(2)$ and a 2-category discrete gamma model. According to Kass and Raftery's scale, the former provides 
a significantly better fit to the data than the latter, which suggests that the substitutional heterogeneity in the data can only be adequately explained by a mixture of $Q$ matrices and confirms that our SS sampler is able to select the true model as the most plausible one.

% (3.2.) Comparison to other marginal likelihood estimators (e.g. HM and AICM?)
\subsection{Comparison to other marginal likelihood estimators}

As a comparison, we conducted model selection using the HM estimator, which can be straightforwardly calculated from the chain of log-likelihoods returned by the MCMC posterior simulation. We simulated $60\,000$ samples from each of the allocation-variable versions of models~\eqref{eq:mixture},~\eqref{eq:mixture_Q} and~\eqref{eq:mixture_r+t} with $k=1,\ldots,4$ components. We thinned the samples to every $10$ iterations and discarded the first quarter as burn-in. For mixtures with four components, we simulated $20\,000$ iterations and thinned and burnt-in the chain in a similar way. We tuned the parameters as for the SS run. We also simulated from a discrete-gamma model using MrBayes with comparable settings and computational effort as for our MCMC runs for HM estimation.

In Figure~\ref{fig:SYNTHETIC_ModelChoice}(b) we have plotted the log-marginal-likelihoods obtained using HM against SS for all considered models. HM estimates exceeded those of SS in all cases, and the pattern is exacerbated as the models become more complex. The difference between the HM and the SS estimates for $Q+t(4)$ is as large as $2\,776$ log units, and almost as large for $Q(4)$. Indeed, HM selects a $Q+t(4)$ mixture even though the data was generated under the simpler $Q+t(2)$ model. The reason is that HM fails to adequately penalise the more complex models for having extra parameters that contribute little to model fit (Xie et al., 2011\nocite{Xieetal_2011}). These results coincide with the growing evidence that HM often overestimates the marginal likelihood making a model appear better-fitting than it really is (Lartillot and Philippe, 2006; Xie et al., 2011; Baele et al., 2012\nocite{Lartillot_Philippe2006,Xieetal_2011,Baeleetal_2012}) and support the notion that HM should be avoided (Calderhead and Girolami, 2009\nocite{Calderhead_Girolami2009}). We do note that our HM and SS estimates are not based on the same number of samples but suggest that the observed patterns in Figure~\ref{fig:SYNTHETIC_ModelChoice}(b) will not be significantly influenced by this.

% (3.3.) Model estimation
\subsection{Model estimation and classification}

\begin{figure} \centering
 {\includegraphics[scale=0.4]{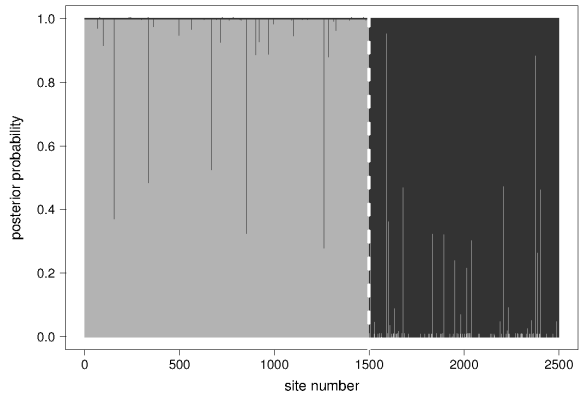}}
\caption{Posterior classification probabilities for the synthetic DNA alignment, from simulation from the posterior of a two-component $Q+t$ mixture model. \usebox{\myblacksquare} and \usebox{\mylightgraysquare} denote the two different mixture components. The dotted line indicates the boundary between the evolutionary classes at generation stage.}
 \label{fig:allocations_synthetic}
\end{figure}

We estimated the $Q+t$ mixture with two components twice, and verified that each of these independent runs converged to the same region in the posterior distribution. The runs for inference comprised $60\,000$ iterations of our MCMC sampler thinned to every $10$ iterations, and we discarded the first quarter as burn-in. The tuning parameters for the proposals remained at the same values as for the SS run. Figure~\ref{fig:allocations_synthetic} shows the estimated posterior probabilities of classification to the two components which, unlike previously published methods (e.g. Pagel and Meade, 2004\nocite{Pagel_Meade2004}; Lartillot and Philippe, 2004\nocite{Lartillot_Philippe2004}; Huelsenbeck and Suchard, 2007\nocite{Huelsenbeck_Suchard2007}), can be directly obtained from the MCMC output as follows. Let $z_{n}^{(1)},\ldots,z_{n}^{(M)}$ be the chain of posterior allocations for site $n$, generated by an MCMC run of length $M$ after burn-in. Variable $z_{n}^{(i)}$ indicates the identity of the component to 
which site $n$ is allocated at iteration $i$ and it takes values in the set $\{1,\ldots,k\}$. Once the chain is checked for convergence to stationarity, good mixing and lack of label-switching, it is used to count the number of times that site $n$ gets allocated to component $j$. This frequency count, divided by the total number of samples, $M$, gives the \emph{posterior classification probability} of site $n$ to component $j$. (Lack of label-switching can be visually verified by inspection of the chain traces; for instance, the $r_{AT}$ and $\pi_{G}$ traces in Supplementary Material III show that the label - or the colour in the case of our visualisation - of each component remains consistent throughout the MCMC run.) In Figure~\ref{fig:allocations_synthetic}, the crossover at which sites were simulated from different evolutionary classes was strikingly well recovered by the method and ergodic posterior averages for the remaining parameters in the mixture coincided favourably with the generating values (Supplementary Material III). The posterior classification probabilities offer a means for not only site classification but also for quantification of classification uncertainty.

\section{Classification of mitochondrial DNA}
\label{sec:analysis_primates}

% (1) Introduce the Primate alignment
\subsection{Methods}
In a second application, we revisited the analysis of mitochondrial DNA (mtDNA) sequences from the primate species human; gorilla; chimpanzee; orangutan; gibbon; crab-eating macaque; common squirrel monkey; Philippine tarsier and ring-tailed lemur (Brown et al., 1982; Hayasaka et al., 1988\nocite{Brown_etal1982, Hayasaka_etal1988}). This alignment, of size $9$ sequences $\times$ $888$ sites after removal of gaps, comprises the portions of two protein-coding genes (sites $1-694$) and a transfer RNA (tRNA) region (sites $695-888$). Transfer RNA is a highly conserved molecule in charge of translating the information encoded by coding genes into the protein alphabet. Such a translation process is achieved by mapping each set of three consecutive, non-overlapping DNA characters within a coding region into one amino acid. A coding DNA triplet is called a codon, and the second position ($cp2$) of a codon is known to undergo substitutions at slower rates than the first ($cp1$) and third codon positions ($cp3$; Fitch and Markowitz, 1970\nocite{Fitch_Markowitz1970}). This difference in substitution rates relates to the fact that a change at the third codon position does not always affect the resulting protein but a change at $cp2$ may, more likely, alter the final product and result in a deleterious mutation.

% (2) Establish its relevance and describe previous analyses of it
In this analysis, we are interested in detecting the evolutionary heterogeneity that exists between the different codon positions and the tRNA region. The primate mtDNA alignment has been analysed extensively using phylogenetic methods (Yang, 1995; Larget and Simon, 1999; Suchard et al., 2001\nocite{Yang1995,Larget_Simon1999, Suchard_etal2001}), in most cases assuming four evolutionary classes (corresponding to the three codon positions plus the tRNA region). Most of these previous approaches have relied on prior knowledge about site membership which may be restrictive and error prone. For instance, in a study by Yang (1995), some sites within the tRNA region were \emph{a priori} misclassified resulting in inaccurate parameter estimates, as stated in the mtprim9.nuc file distributed with the software package PAML4 (Yang, 2007\nocite{Yang2007}).

% (3) Describe the results
% (3.1.) Model selection
\subsection{Model selection}

\begin{figure} \centering
{\includegraphics[scale=0.45]{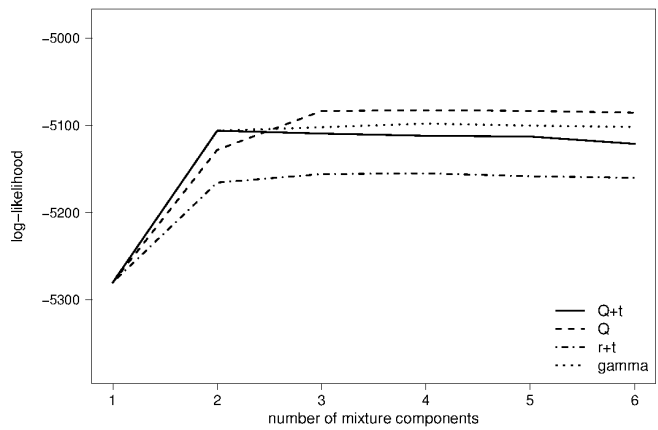}}
\caption{Estimated log marginal likelihoods for the models fitted to the primate mtDNA alignment using the SS sampler. Plotted data: homogeneous: $-5\,280.07$; $Q+t$ with 2--6 components: $-5\,106.11$, $-5\,109.42$, $-5\,111.94$, $-5\,112.85$, $-5\,121.21$; $Q$ with 2--6 components: $-5\,128.15$, $-5\,083.56$, $-5\,082.61$, $-5\,083.46$, $-5\,085.46$; $r+t$ with 2--6 components: $-5\,165.62$, $-5\,155.88$, $-5\,155.02$, $-5\,158.40$, $-5\,160.15$; gamma with 2--6 categories: $-5\,106.20$, $-5\,102.17$, $-5\,097.97$, $-5\,100.35$, $-5\,101.87$.}
 \label{fig:PRIMATES_ModelChoice}
\end{figure}

We considered $Q+t$, $Q$ and $r+t$ mixtures, with different number of components, for the primate mtDNA alignment. Figure~\ref{fig:PRIMATES_ModelChoice} shows the log marginal likelihoods of these models, estimated using the SS sampler. The proposals were tuned to $\delta=1.5$ for the BLM move; $\sigma=0.06$ for BLNA; $\alpha_r=800$, $\alpha_{\pi}=600$ and $\epsilon_{\theta}=0.0001$ for the $\epsilon$Dirichlet proposal for substitution rates and stationary probabilities; and $\epsilon_{\omega}=0.0001$ for mixture proportions. Hyperparameter $\eta$ for the prior on a branch length was set to $2.5$, in line with Suchard et al. (2001\nocite{Suchard_etal2001}). Following a burn-in phase consisting of $20\,000$ iterations at power $\beta=1$ (the posterior), the SS sampler was run for $5\,000$ iterations for each $\beta$ value in the steppingstone path ($K=30$ steps in total). 

For comparison, we fitted the data with a discrete-gamma model using the SS sampler in MrBayes 3.2. In Figure~\ref{fig:PRIMATES_ModelChoice} it is clear that the data contain heterogeneity that is not fully accounted for by either the homogeneous or the discrete-gamma models. A $Q$ mixture with three components improved upon the homogeneous model by nearly $197$ log-units, and upon the discrete-gamma model with $2$ and $3$ categories by approximately $23$ and $19$ log-units, respectively. In all cases, the Kass and Raftery (1995\nocite{Kass_Raftery1995}) scale indicated very strong evidence in favour of a $Q(3)$ mixture. A four-component $Q$ mixture continued to improve upon the $Q$ mixture with three components, but this improvement was non-significant, i.e. $2ln(\mbox{BF}_{Q(4)\,vs\,Q(3)})<2$. The model choice mechanism, therefore, pointed towards the $Q(3)$ mixture as the most plausible model for the data.

% I am reporting log-MLs to two decimal places because of the comparison Q(4) vs. Q(3) that I make in the text.

% (3.2.) Model estimation
\subsection{Model estimation and classification}
\label{sec:Primate_model_estimation}

Two independent runs for inference were conducted comprising $40\,000$ iterations of our MCMC algorithm, with no thinning, preceded by $15\,000$ cycles as burn in. Examination of trace plots of the log-likelihood, the observed consistency between runs and our experience with the SS runs suggested that the burn-in period was sufficiently long. We also confirmed that the runs did not suffer from label-switching by examination of the MCMC trace plots. The tuning parameters for the proposals remained at the same values as for the SS run. 

\begin{figure} \centering
 {\includegraphics[scale=0.45]{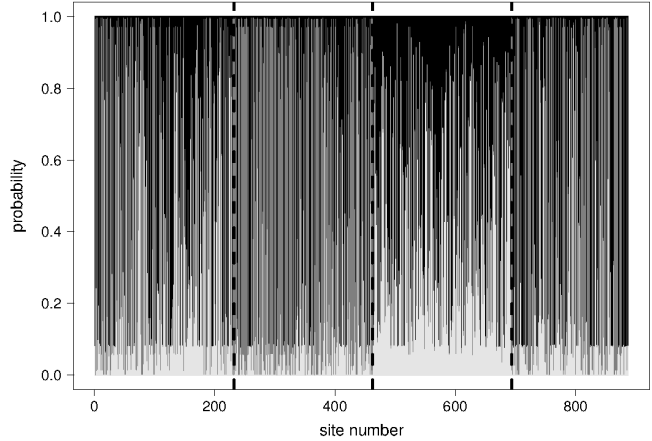}}
\caption{Posterior classification probabilities for the primate mtDNA alignment, from an analysis with a three-component $Q$ mixture. Classification probabilities to each component are differentiated by colour: \usebox{\myblacksquare}, \usebox{\mydarkgraysquare} and \usebox{\mylightgraysquare} denote the three different components. The dotted lines separate the regions $cp1\,\mid\,cp2\,\mid\,cp3\,\mid\,\mbox{tRNA}$ in the alignment.}
 \label{fig:allocation_mtDNA}
\end{figure}

Figure~\ref{fig:allocation_mtDNA} shows the estimated posterior classification probabilities of sites belonging to each of the three components in the mixture. For ease of visual interpretation, the protein-coding genes have been rearranged according to codon position; sites $1-232$ correspond to $cp1$, sites $233-463$ to $cp2$ and sites $464-694$ to $cp3$, but there is nothing in the formulation of the classification method that requires such a rearrangement. Two clear patterns emerged: sites in the highly conserved $cp2$ and tRNA regions were mostly allocated to component~\usebox{\mydarkgraysquare}, whereas the $cp3$ region is clearly dominated by components~\usebox{\myblacksquare} and~\usebox{\mylightgraysquare}. The method is able to capture the qualitatively different patterns of evolution in the data without prior partition into classes: the $cp1$ and $cp3$ regions are evolutionarily distinct to the $cp2$ and tRNA classes, with the bulk of this difference being observed between the $cp3$ and the $cp2\,/\,\mbox{tRNA}$ classes. Moreover, our approach allows for probabilistic statements of site classification such as \emph{\textquotedblleft site 1 belongs to component~\usebox{\myblacksquare} with probability $0.91690$, to component~\usebox{\mydarkgraysquare} with probability $0.00007$ and to component~\usebox{\mylightgraysquare} with probability $0.08302$\textquotedblright}.

Our approach allows inferences on evolutionary model parameters for individual mixture components and branch lengths. Table~\ref{table:mtDNA_estimates} reports the ergodic average of parameters for each component distribution. Components~\usebox{\myblacksquare} and~\usebox{\mylightgraysquare}, which seem to specialise in $cp3$, show ergodic averages for the rates of substitution that agree with the bias that favours transitions (a substitution from $A\rightarrow G$ or $C\rightarrow T$) over transversions (any other substitution). This transition bias has been reported before (Wakeley, 1996\nocite{Wakeley1996}) and one of its suggested causes is the fact that only about 3\% of transitions at the third codon position cause amino acid changes compared with 41\% of transversions (Crozier and Crozier, 1993\nocite{Crozier_Crozier1993}). Component \usebox{\mydarkgraysquare} shows extremely low $A$ content, which clearly leads to poor $r_{AC},\,r_{AG}$ and $r_{AT}$ estimates (Supplementary Material IV). This illustrates that the estimation of some parameters on an individual component basis may be discouraged for some applications for which the component-specific data do not contain the signal required to estimate all phylogenetic parameters. In such cases, one could specify mixtures of evolutionary models with constrained cases of the GTR model (Evans and Sullivan, 2012\nocite{Evans_Sullivan2012}; not attempted here) or restrict inference to the allocation variables, $z_1,\ldots,z_N$. 
% 
% \begin{table}[t]
%  \begin{center}
%  \begin{tabular*}{1\textwidth}{@{\extracolsep{\fill}} c | c r | c r | c r }
%  \multirow{2}{0.4cm}{} & \multicolumn{2}{c|}{component \usebox{\myblacksquare}} & \multicolumn{2}{c|}{component \usebox{\mydarkgraysquare}} & \multicolumn{2}{c}{component \usebox{\mylightgraysquare}}\\
% 		& \scriptsize \itshape ergodic average & $\hat{\tau}$ & \scriptsize \itshape ergodic average & $\hat{\tau}$ & \scriptsize \itshape ergodic average & $\hat{\tau}$ \\
%  \hline
%  $r_{AC}$ & 0.0048 & 184 & 0.0451 & 110 & 0.0893 & 932 \\
%  $r_{AG}$ & 0.4325 & 413 & 0.3981 & 1179 & 0.2500 & 831 \\
%  $r_{AT}$ & 0.0112 & 129 & 0.5102 & 192 & 0.1309 & 1500 \\
%  $r_{CG}$ & 0.0608 & 126 & 0.0039 & 377 & 0.0352 & 100 \\
%  $r_{CT}$ & 0.4173 & 519 & 0.0414 & 521 & 0.4535 & 396 \\
%  $r_{GT}$ & 0.0734 & 157 & 0.0013 & 488 & 0.0411 & 69 \\
%  $\pi_{A}$ & 0.6022 & 174 & 0.0088 & 256 & 0.4269 & 61 \\
%  $\pi_{C}$ & 0.2075 & 150 & 0.3314 & 87 & 0.3257 & 40 \\
%  $\pi_{G}$ & 0.0259 & 351 & 0.1913 & 120 & 0.1221 & 71 \\
%  $\pi_{T}$ & 0.1644 & 347 & 0.4685 & 499 & 0.1253 & 51 \\
%  $\omega_j$ & 0.3765 & 279 & 0.3858 & 261 & 0.2377 & 66 \\
%  \end{tabular*}
% \caption{Ergodic averages of model parameters and estimated integrated autocorrelation times, $\hat{\tau}$, from an analysis of the primate mtDNA alignment with a three-component $Q$ mixture.}
% \label{table:mtDNA_estimates}
%  \end{center}
% \end{table}

\begin{table}[t]
 \begin{center}
 \rowcolors{2}{}{light-light-gray}
 \begin{tabular}{ c c c c }
 & component \usebox{\myblacksquare} & component \usebox{\mydarkgraysquare} & component \usebox{\mylightgraysquare}\\
 \hline
 $r_{AC}$ & 0.0048 & 0.0451 & 0.0893\\
 $r_{AG}$ & 0.4325 & 0.3981 & 0.2500\\
 $r_{AT}$ & 0.0112 & 0.5102 & 0.1309\\
 $r_{CG}$ & 0.0608 & 0.0039 & 0.0352\\
 $r_{CT}$ & 0.4173 & 0.0414 & 0.4535\\
 $r_{GT}$ & 0.0734 & 0.0013 & 0.0411\\
 $\pi_{A}$ & 0.6022 & 0.0088 & 0.4269\\
 $\pi_{C}$ & 0.2075 & 0.3314 & 0.3257\\
 $\pi_{G}$ & 0.0259 & 0.1913 & 0.1221\\
 $\pi_{T}$ & 0.1644 & 0.4685 & 0.1253\\
 $\omega_j$ & 0.3765 & 0.3858 & 0.2377\\
 \end{tabular}
\caption{Ergodic averages of model parameters from an analysis of the primate mtDNA alignment with a three-component $Q$ mixture. For a visual measure of parameter uncertainty, refer to Supplementary Material IV.}
\label{table:mtDNA_estimates}
 \end{center}
\end{table}

\begin{figure}[tb] \centering
{\includegraphics[scale=0.5]{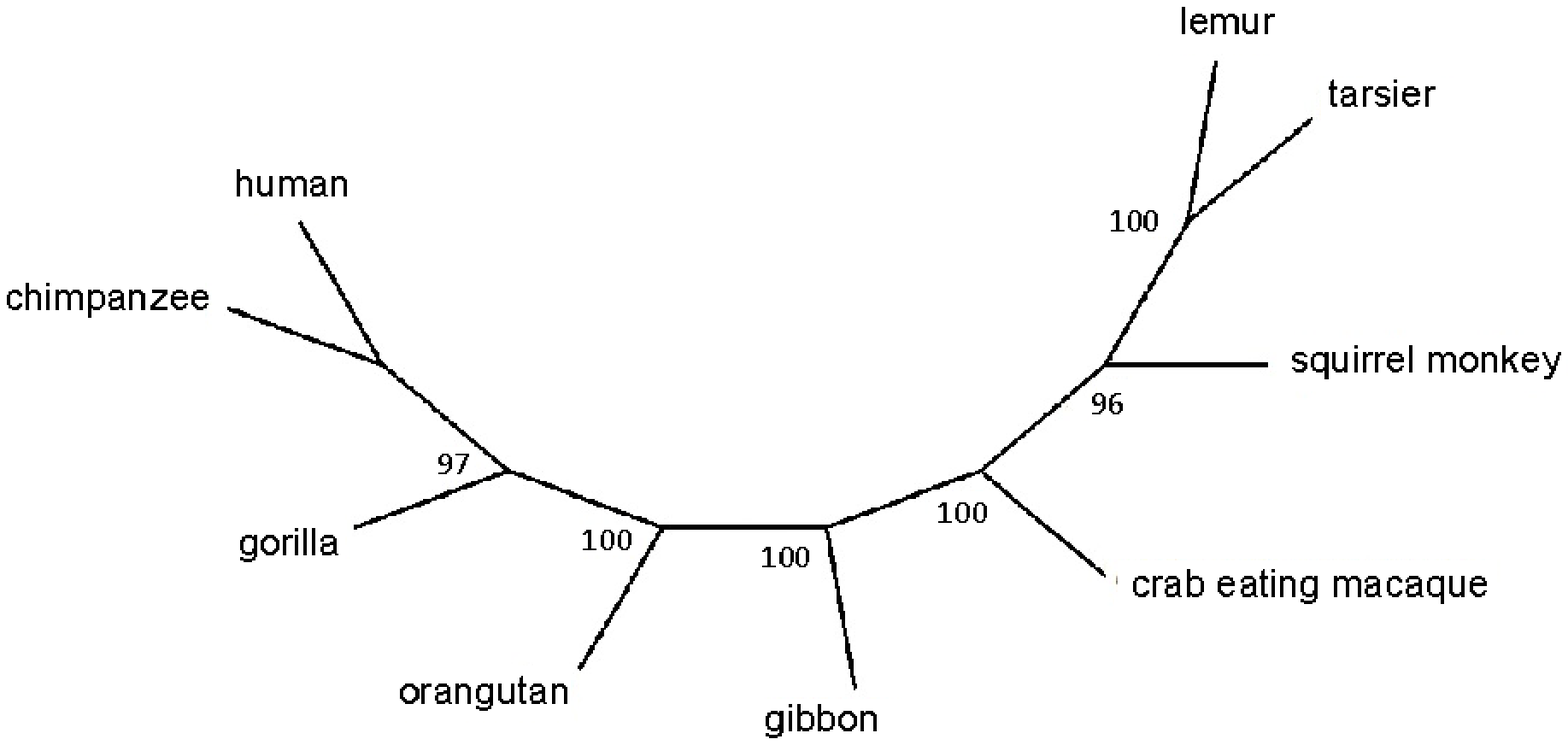}}
\caption{The 50\% majority-rule consensus topology obtained from the chain of sampled topologies during the analysis of the primate mtDNA alignment with a three-component $Q$ mixture. The numbers at the nodes indicate the percent of topologies, within the chain of sampled topologies, in which that clade is present.}
 \label{fig:PRIMATES_ConsensusTop}
\end{figure}

The consensus tree topology, obtained as the 50\% majority-rule, is shown in Figure~\ref{fig:PRIMATES_ConsensusTop}. This topology agrees favourably with the published topologies in Yang (1995), Larget and Simon (1999) and Suchard et al. (2001)\nocite{Yang1995,Larget_Simon1999,Suchard_etal2001}. The total length of interior and exterior branches, estimated as the ergodic average of post-burn-in samples, was $0.8574$ and $2.4410$, respectively (estimated uncertainties are visually reported in Supplementary Material IV). 

A mixture model augmented with allocation variables was successful in describing the heterogeneity present in the primate mtDNA alignment. The analysis further allowed us to visualise underlying structural information. Regions in the alignment that are known to be highly conserved ($cp2$ and $tRNA$) were grouped in a common component whereas the highly variable $cp3$ region was classified to a distinct component. Such a structure discovery process could be applied to molecular sequence data for which \emph{a priori} partition information is not available. For instance, at the moment of writing, the leading author of this paper is using phylogenetic mixture models to classify hundreds of genes into groups. The ultimate goal is to select only a few representatives per group and thus reduce the dimensionality of the problem, i.e. to move from a problem that includes hundreds of genes to one that only deals with a few.

%(3.4)
\subsection{Predictive uses for allocation variables}
\label{sec:allocations}

The allocation variables form a coherent basis for site classification. In addition to providing the means to classify the observed sites via the posterior probabilities $\{p(z_n=j\mid x); j=1,2,\ldots,k\}$, they can also be used to classify sites on a predictive basis. To illustrate this we address the question of classifying a future (unobserved) site pattern $x^{*}_{n}$. If the allocation variable corresponding to such a site pattern is $z^{*}_{n}$ and assuming $k$ is known, then we are interested in the probability

\begin{equation}
  p(z^{*}_{n}=j\mid x,x^{*}_{n}) \nonumber
\end{equation}

\noindent which can be approximated as (Richardson and Green, 1997\nocite{Richardson_Green1997}):

\begin{equation}
 \begin{split}
   p(z^{*}_{n}=j\mid x,x^{*}_{n}) &= \int p(z^{*}_{n}=j\mid x, x^{*}_{n}, \phi, t, \theta, \omega) \,p(\phi, t, \theta, \omega \mid x, x^{*}_{n}) \,d\phi \,dt \,d\theta \,d\omega \\
                                  &= \int p(z^{*}_{n}=j\mid x^{*}_{n}, \phi, t, \theta, \omega) \,p(\phi, t, \theta, \omega \mid x, x^{*}_{n}) \,d\phi \,dt \,d\theta \,d\omega\\
                                  &\thickapprox \int p(z^{*}_{n}=j\mid x^{*}_{n}, \phi, t, \theta, \omega) \,p(\phi, t, \theta, \omega \mid x) \,d\phi \,dt \,d\theta \,d\omega \nonumber
 \end{split}
\end{equation}

\noindent and the last integral is estimated by the MCMC empirical average

\begin{equation}
  \sum_{i=1}^{M} \left[\omega_j\,p(x^{*}_{n}\mid \phi, t, \theta_j)\,/\,\sum_{j=1}^k \omega_j\,p(x^{*}_{n}\mid \phi, t, \theta_j)\right]_i\,/\,M 
\label{eq:pred_probability}
\end{equation}

\noindent where $p(\centerdot \mid \phi, t, \theta_j)$ is the $j$th component distribution of a $Q$ mixture with $k$ components and $M$ is the number of MCMC samples after burn-in. Table~\ref{table:predictive_probs} shows the predictive classification probabilities of $16$ different site patterns, of the many tens of thousands we could consider, calculated using equation~\eqref{eq:pred_probability} and the MCMC sample generated when the $Q(3)$ mixture was fitted to the primate mtDNA alignment (Section~\ref{sec:Primate_model_estimation}). All of these site patterns are unobserved in the primate alignment.

\begin{table}[!ht]
    \begin{center}
	\setlength\extrarowheight{4pt}
        \rowcolors{2}{}{light-light-gray}
	\begin{tabular}{l c l c l c l}
            \small Pattern & \small $p(z^{*}_{n}=\usebox{\myblacksquare}\mid \centerdot)$ &  & \small $p(z^{*}_{n}=\usebox{\mydarkgraysquare}\mid \centerdot)$ &  & \small $p(z^{*}_{n}=\usebox{\mylightgraysquare}\mid \centerdot)$ & \\
            \hline
$h$=C, $r$=A        & 0.2682 & & 0.0000 & & 0.7318 &\\
$h$=T, $r$=A        & 0.5125 & & 0.0000 & & 0.4875 &\\
$h$=A, $r$=C        & 0.1069 & & 0.0635 & & 0.8296 &\\
$h$=G, $r$=C        & 0.1721 & & 0.4610 & & 0.3669 &\\
$h$=T, $r$=C        & 0.2875 & & 0.5097 & & 0.2028 &\\
$h$=A, $r$=G       & 0.0049 & & 0.5320 & & 0.4632 &\\
$h$=C, $r$=G       & 0.0016 & & 0.8118 & & 0.1866 &\\
$h$=T, $r$=G       & 0.0009 & & 0.9176 & & 0.0814 &\\
$h$=A, $r$=T       & 0.0585 & & 0.8543 & & 0.0872 &\\
$h$=C, $r$=T       & 0.1750 & & 0.7638 & & 0.0612 &\\
$h$=G, $r$=T       & 0.0909 & & 0.8601 & & 0.0490 &\\[0.1cm]

% (o) A, (r) A & 0.8001 & 0.0000 & 0.1999\\
% (o) C, (r) A & 0.2876 & 0.0000 & 0.7124\\
% (o) G, (r) A & 0.7437 & 0.0000 & 0.2563\\
% (o) T, (r) A & 0.5226 & 0.0000 & 0.4774\\
% (o) A, (r) C & 0.1192 & 0.0565 & 0.8243\\
% (o) C, (r) C & 0.1872 & 0.4735 & 0.3393\\
% (o) G, (r) C & 0.1574 & 0.4710 & 0.3716\\
% (o) T, (r) C & 0.2837 & 0.5185 & 0.1978
        \end{tabular}
\caption{Predictive classification probabilities of sites with patterns as indicated in the first column, where $h$ is the character in the human sequence and $r$ is the characters in the rest of the sequences.}
    \label{table:predictive_probs}
    \end{center}
\end{table}

This exercise illustrates the potential of allocation variables as tools for classification of site patterns that are not observed in an alignment. Given an alignment of multiple molecular sequences and an MCMC run that estimates a given mixture model such as the ones presented in this paper (although not restricted to them), one could ask what is the probability of an unobserved site pattern belonging to each mixture component. This can be of direct relevance in genomic studies, where it is impractical to rerun the MCMC sampler if a new allele is sequenced; one could instead simply compute the predictive classification probabilities of those site patterns that are novel in the newly sequenced allele. 

% 
% I've commented these out as perhaps they answer a question the reviewer asks but not one which the potential reader of this paper would ask - possibly they would sit better in the "Response"?
%The allocation variables are also of interest from a theoretical point of view. In the MCMC implementation, samples for the allocation variables $z$ and the mixture parameters $(\omega,\phi,t,\theta)$ are alternately generated, producing an allocation-variable chain and a parameter chain, both of which are Markov. The finite-state structure of the allocation-variable chain allows convergence results, such as geometric convergence, $\phi$-mixing and a central limit theorem, to be easily established for it and transferred automatically to the parameter chain (Robert, 1996\nocite{Robert1996} and references therein). We have not used the allocation-variable chain for such purposes in this work but simply emphasise their utility to the reader.

\section{Discussion}

We have presented a method that employs phylogenetic mixture models augmented with allocation variables to produce site classification estimates and a measure of their uncertainty. 

The mixture models that we have chosen for demonstrating site classification account for both qualitative and quantitative among-site rate variation by including multiple sets of branch lengths and $Q$ matrices. Mixtures with multiple sets of branch lengths account for a phenomenon known as heterotachy (Lopez, Casane and Philippe, 2002\nocite{Lopez_etal2002}), in which the rates of evolution along the branches leading to different taxa in the tree vary across sites. Since the beginning of this research a number of groups have independently proposed mixtures of sets of branch lengths as a way of modelling heterotachy in phylogenetic studies (Pagel and Meade, 2008\nocite{Pagel_Meade2008}). We do note, however, that models $Q+t$, $Q$ and $r+t$ may not be suitable in every situation. For heterotachy-free data, mixture models such as those proposed in Evans and Sullivan (2012)\nocite{Evans_Sullivan2012} might suffice (i.e. mixtures of multiple evolutionary models, multiple scaling factors of these evolutionary models but only one set of branch lengths). Or, constraining Evans and Sullivan's models one level further, the scaling factors could be made to conform to a discrete gamma distribution with empirically estimated shape, i.e., a generalisation of the discrete-gamma model of Yang (1994\nocite{Yang1994a}). This would result in less parameter-rich models than the ones presented in this work that may, or may not, provide a more plausible fit to the data in question. In problems where molecular data are suspected to have undergone recombination, a fit with either of $Q+t$, $Q$ or $r+t$ could be inadequate because contiguous subsequences of recombinant bacterial or viral DNA would be expected to follow different phylogenetic histories; a model with class-specific topologies would then be required (Kitchen et al., 2009\nocite{Kitchen_etal2009}). Whatever the flavour of the phylogenetic mixture, we hope that the main message of our paper is clear: Bayesian mixture models can be extended to include allocation variables and be readily used as tools for site classification. The reader is encouraged to always follow the principles of a valid inferential process by considering a set of candidate models and conducting model selection before estimating a model (Fisher, 1922\nocite{Fisher1922}). It is clear that more flexible and user-friendly software tools to conduct such model selection processes in phylogenetics are required.

A potential application of our classification method, not pursued here, is as a tool for identifying the sites that are unable to undergo substitution. The presence of invariant sites is a well-documented cause of inconsistency in phylogeny reconstruction (Steel et al., 2000\nocite{Steeletal_2000}), and \textquoteleft augmented\textquoteright\, phylogenetic mixtures could be deployed to pinpoint the invariant sites that should be excluded from the alignment before inference. This idea has been discussed elsewhere (Huelsenbeck and Suchard, 2007\nocite{Huelsenbeck_Suchard2007}) and its implementation would require defining a mixture that includes a strictly invariant class (i.e. a component in which all the rates of substitution are zero).

We recognise the limitations of the NNI proposal in our MCMC sampler and note that larger problems may require additional topology proposal mechanisms. The topology update mechanism that our models require is restricted to updating the tree topology while preserving the branch lengths due to the fact that there is only one topology shared across all components. The applications that we present in this paper are of modest size and we are confident that NNI successfully leads the MCMC chain into stationarity; visual inspection of the post-burn-in log-likelihood trace of the primate mtDNA alignment indicate lack of pre-stationary trends and our 50\% majority-rule consensus tree agrees favourably with several other published studies. Tackling larger problems may require additional MCMC mechanisms and this is regarded as an aspect for future work.



\section*{Acknowledgements}

Elisa Loza was supported by the Mexican Council for Science and Technology (CONACYT) during the production of the research and partially supported by Rothamsted Research during the preparation of the manuscript. Rothamsted Research receives grant-aided support from the Biotechnology and Biological Sciences Research Council (BBSRC) of the UK. Merrilee Hurn and Tony Robinson were supported by the Bath Institute for Complex Systems (EPSRC grant GR/S86525/01). We thank Klaus Kurtenbach, Gabi Margos and Ed Feil, all from the Department of Biology and Biochemistry at the University of Bath, for their helpful discussions. Thanks also to Kevin Dawson, from the Wellcome Trust Sanger Institute, for valuable comments to earlier versions of this manuscript. We thank two anonymous reviewers for providing valuable criticism that improved the quality of this paper.

\clearpage
\bibliographystyle{acm}
\bibliography{Loza-Reyes_etal2013}

\begin{thebibliography}{10}

\bibitem{Ardia_etal2012}
{\sc Ardia, D., Ba\c{s}t\"{u}rk, N., Hoogerheide, L., and van Dijk, H.~K.}
\newblock {A comparative study of Monte Carlo methods for efficient evaluation
  of marginal likelihood}.
\newblock {\em {Computational Statistics and Data Analysis} 56\/} (2012),
  3398--3414.

\bibitem{Baeleetal_2012}
{\sc Baele, G., Lemey, P., Bedford, T., Rambaut, A., Suchard, M.~A., and
  Alekseyenko, A.~V.}
\newblock {Improving the Accuracy of Demographic and Molecular Clock Model
  Comparison While Accommodating Phylogenetic Uncertainty}.
\newblock {\em Molecular Biology and Evolution 29}, 9 (2012), 2157--2167.

\bibitem{Brown_etal1982}
{\sc Brown, W.~M., Prager, E.~M., Wang, A., and Wilson, A.~C.}
\newblock {Mitochondrial DNA sequences of primates: tempo and mode of
  evolution}.
\newblock {\em Journal of Molecular Evolution 18\/} (1982), 225--239.

\bibitem{Calderhead_Girolami2009}
{\sc Calderhead, B., and Girolami, M.}
\newblock {Estimating Bayes factors via thermodynamic integration and
  population MCMC}.
\newblock {\em {Computational Statistics and Data Analysis} 53\/} (2009),
  4028--4045.

\bibitem{Crozier_Crozier1993}
{\sc Crozier, R.~H., and Crozier, Y.~C.}
\newblock {The mitochondrial genome of the honeybee \textit{Apis mellifera}:
  complete sequence and genome organization}.
\newblock {\em Genetics 133\/} (1993), 97--117.

\bibitem{Evans_Sullivan2012}
{\sc Evans, J., and Sullivan, J.}
\newblock {Generalized Mixture Models for Molecular Phylogenetic Estimation}.
\newblock {\em {Systematic Biology} 61}, 1 (2012), 12--21.

\bibitem{Fanetal_2011}
{\sc Fan, Y., Wu, R., Chen, M.-H., Kuo, L., and Lewis, P.~O.}
\newblock {Choosing among Partition Models in Bayesian Phylogenetics}.
\newblock {\em Molecular Biology and Evolution 28}, 1 (2011), 523--432.

\bibitem{Felsenstein1981}
{\sc Felsenstein, J.}
\newblock Evolutionary trees from {DNA} sequences: {A} maximum likelihood
  approach.
\newblock {\em Journal of Molecular Evolution 17\/} (1981), 368--376.

\bibitem{Felsenstein_Churchill1996}
{\sc Felsenstein, J., and Churchill, G.~A.}
\newblock A hidden {Markov} model approach to variation among sites in rate of
  evolution.
\newblock {\em Molecular Biology and Evolution 13}, 1 (1996), 93--104.

\bibitem{Fisher1922}
{\sc Fisher, R.~A.}
\newblock {On the mathematical foundations of theoretical statistics}.
\newblock {\em {Philosophical Transactions of the Royal Society (Series A)}
  222\/} (1922), 309--368.

\bibitem{Fitch_Markowitz1970}
{\sc Fitch, W.~M., and Markowitz, E.}
\newblock An improved method for determining codon variability in a gene and
  its application to the rate of fixation of mutations in evolution.
\newblock {\em Biochemical Genetics 4\/} (1970), 579--593.

\bibitem{friel2012improving}
{\sc Friel, N., Hurn, M., and Wyse, J.}
\newblock Improving power posterior estimation of statistical evidence.
\newblock To appear in Statistics and Computing, DOI:
  10.1007/s11222-013-9397-1, 2013.

\bibitem{friel2008marginal}
{\sc Friel, N., and Pettitt, A.}
\newblock Marginal likelihood estimation via power posteriors.
\newblock {\em Journal of the Royal Statistical Society B 70}, 3 (2008),
  589--607.

\bibitem{green1995reversible}
{\sc Green, P.}
\newblock Reversible jump {M}arkov chain {M}onte {C}arlo computation and
  {B}ayesian model determination.
\newblock {\em Biometrika 82}, 4 (1995), 711--732.

\bibitem{Hasegawa_etal1985}
{\sc Hasegawa, M., Kishino, H., and Yano, T.}
\newblock Dating of the human-ape splitting by a molecular clock of
  mitochondrial {DNA}.
\newblock {\em Journal of Molecular Evolution 22\/} (1985), 160--174.

\bibitem{Hayasaka_etal1988}
{\sc Hayasaka, K., Gojobori, T., and Horai, S.}
\newblock {Molecular phylogeny and evolution of primate mitochondrial DNA}.
\newblock {\em Molecular Biology and Evolution 5}, 6 (1988), 626--644.

\bibitem{Huelsenbeck_Suchard2007}
{\sc Huelsenbeck, J., and Suchard, M.}
\newblock {A Nonparametric Method for Accommodating and Testing Across-Site
  Rate Variation}.
\newblock {\em {Systematic Biology} 56}, 6 (2007), 975--987.

\bibitem{Huelsenbeck_Ronquist2001}
{\sc Huelsenbeck, J.~P., and Ronquist, F.}
\newblock {MrBayes: Bayesian inference of phylogeny}.
\newblock {\em Bioinformatics 17\/} (2001), 754--755.

\bibitem{Hurn08}
{\sc Hurn, M.~A., Green, P.~J., and Al-Awadhi, F.}
\newblock A {B}ayesian hierarchical model for photometric redshifts.
\newblock {\em Applied Statistics, Journal of the Royal Statistical Society C
  57\/} (2008), 487--504.

\bibitem{Jasra_Holmes_Stephens2005}
{\sc Jasra, A., Holmes, C.~C., and Stephens, D.~A.}
\newblock {Markov chain Monte Carlo methods and the label switching problem in
  Bayesian mixture modeling}.
\newblock {\em Statistical Science 20}, 1 (2005), 50--67.

\bibitem{Jukes_Cantor1969}
{\sc Jukes, T., and Cantor, C.}
\newblock {Evolution of protein molecules}.
\newblock In {\em {Mammalian Protein Metabolism}}, H.~N. Munro, Ed. Academic
  Press, New York, USA, 1969, pp.~21--132.

\bibitem{Kass_Raftery1995}
{\sc Kass, R.~E., and Raftery, A.~E.}
\newblock {Bayes Factors}.
\newblock {\em Journal of the American Statistical Association 90}, 430 (1995),
  773--795.

\bibitem{Kitchen_etal2009}
{\sc Kitchen, C. M.~R., Kroll, J., Kuritzkes, D.~R., Bloomquist, E., Deeks,
  S.~G., and Suchard, M.~A.}
\newblock {Two-way Bayesian hierarchical phylogenetic models: An application to
  the co-evolution of gp120 and gp41 during and after enfuvirtide treatment}.
\newblock {\em {Computational Statistics and Data Analysis} 53\/} (2009),
  766--775.

\bibitem{Kolaczkowski_Thornton2008}
{\sc Kolaczkowski, B., and Thornton, J.~W.}
\newblock {A mixed branch length model of heterotachy improves phylogenetic
  accuracy}.
\newblock {\em {Molecular Biology and Evolution} {25}}, {6} ({JUN} {2008}),
  {1054--1066}.

\bibitem{Lanave_etal1984}
{\sc Lanave, C., Preparata, G., Saccone, C., and Serio, G.}
\newblock {A new method for calculating evolutionary substitution rates}.
\newblock {\em Journal of Molecular Evolution 20\/} (1984), 86--93.

\bibitem{Larget_Simon1999}
{\sc Larget, B., and Simon, D.~L.}
\newblock {Markov chain Monte Carlo algorithms for the Bayesian analysis of
  phylogenetic trees}.
\newblock {\em Molecular Biology and Evolution 16}, 6 (1999), 750--759.

\bibitem{Lartillot_Philippe2004}
{\sc Lartillot, N., and Philippe, H.}
\newblock {A Bayesian Mixture Model for Across-Site Heterogeneities in the
  Amino-Acid Replacement Process}.
\newblock {\em Molecular Biology and Evolution 21}, 6 (2004), 1095--1109.

\bibitem{Lartillot_Philippe2006}
{\sc Lartillot, N., and Philippe, H.}
\newblock {Computing Bayes factors using thermodynamic integration}.
\newblock {\em {Systematic Biology} {55}\/} ({2006}), {195--207}.

\bibitem{Leslie07}
{\sc Leslie, D.~S.}
\newblock {Discussion on Model-based clustering for social networks (by M. S.
  Handcock, A. E. Raftery and J. M. Tantrum)}.
\newblock {\em Journal of the Royal Statistical Society A 170\/} (2007),
  301--354.

\bibitem{Lopez_etal2002}
{\sc Lopez, P., Casane, D., and Philippe, H.}
\newblock {Heterotachy, an important process of protein evolution}.
\newblock {\em {Molecular Biology and Evolution} {19}}, {1} ({JAN} {2002}),
  {1--7}.

\bibitem{Meade_Pagel2008}
{\sc Meade, A., and Pagel, M.}
\newblock {A Phylogenetic Mixture Model for Heterotachy}.
\newblock In {\em {Evolutionary Biology from Concept to Application}},
  P.~Pontarotti, Ed., first~ed. Springer-Verlag, 2008, pp.~29--41.

\bibitem{Moore_etal1973}
{\sc Moore, G.~W., Goodman, M., and Barnabas, J.}
\newblock An iterative approach from the stand-point of the additive hypothesis
  to the dendrogram problem posed by molecular data sets.
\newblock {\em Journal of Theoretical Biology 38\/} (1973), 423--457.

\bibitem{Newton_Raftery1994}
{\sc Newton, M., and Raftery, A.}
\newblock {Approximate Bayesian inference by the weighted likelihood
  bootstrap}.
\newblock {\em {Journal of the Royal Statistical Society Series B} 56\/}
  (1994), 3--48.

\bibitem{Pagel_Meade2004}
{\sc Pagel, M., and Meade, A.}
\newblock {A phylogenetic mixture model for detecting pattern-heterogeneity in
  gene sequence or character-state data}.
\newblock {\em Systematic Biology 56}, 4 (2004), 571--581.

\bibitem{Pagel_Meade2008}
{\sc Pagel, M., and Meade, A.}
\newblock {Modelling heterotachy in phylogenetic inference by reversible-jump
  Markov chain Monte Carlo}.
\newblock {\em Philosophical Transactions of the Royal Society B 363\/} (2008),
  3955--3964.

\bibitem{Rambaut_Grassly1997}
{\sc Rambaut, A., and Grassly, N.~C.}
\newblock {Seq-Gen: An application for the Monte Carlo simulation of DNA
  sequence evolution along phylogenetic trees}.
\newblock {\em Computer Applications in the Biosciences 13\/} (1997), 235--238.

\bibitem{Richardson_Green1997}
{\sc Richardson, S., and Green, P.~J.}
\newblock {On Bayesian analysis of mixtures with an unknown number of
  components}.
\newblock {\em {Journal of the Royal Statistical Society B} {59}}, {4}
  ({1997}), {731--758}.

\bibitem{Robinson1971}
{\sc Robinson, D.~F.}
\newblock {Comparison of labeled trees with valency three}.
\newblock {\em Journal of Combinatorial Theory 11\/} (1971), 105--119.

\bibitem{Ronquist_etal2012}
{\sc Ronquist, F., Teslenko, M., van~der Mark, P., Ayres, D.~L., Darling, A.,
  Hohna, S., Larget, B., Liu, L., Suchard, M.~A., and Huelsenbeck, J.~P.}
\newblock {MrBayes 3.2: Efficient Bayesian Phylogenetic Inference and Model
  Choice Across a Large Model Space}.
\newblock {\em {Systematic Biology} {61}}, {3} ({2012}), {539--542}.

\bibitem{Steel_2005}
{\sc Steel, M.}
\newblock {Should phylogenetic models be trying to "fit an elephant"}.
\newblock {\em Trends in Genetics 21}, 6 (2005), 307--309.

\bibitem{Steeletal_2000}
{\sc Steel, M., Huson, D., and Lockhart, P.~J.}
\newblock {Invariable Sites Models and Their Use in Phylogeny Reconstruction}.
\newblock {\em Systematic Biology 49\/} (2000), 225--232.

\bibitem{Suchard_etal2001}
{\sc Suchard, M.~A., Weiss, R.~E., and Sinsheimer, J.~S.}
\newblock Bayesian selection of continuous-time {Markov} chain evolutionary
  models.
\newblock {\em Molecular Biology and Evolution 18}, 6 (2001), 1001--1013.

\bibitem{Tavare1986}
{\sc Tavar\'{e}, S.}
\newblock {Some probabilistic and statistical problems in the analysis of DNA
  sequences}.
\newblock In {\em {Lectures on Mathematics in the Life Sciences}}, R.~M. Miura,
  Ed., vol.~17. American Mathematical Society, Providence, USA, 1986,
  pp.~57--86.

\bibitem{Verdinelli_Wasserman1995}
{\sc Verdinelli, I., and Wasserman, L.}
\newblock {Computing Bayes factors using a generalization of the Savage-Dickey
  density ratio}.
\newblock {\em Journal of the American Statistical Association 90\/} (1995),
  614--618.

\bibitem{Wakeley1996}
{\sc Wakeley, J.}
\newblock The excess of transitions among nucleotide substitutions: new methods
  of estimating transition bias underscore its significance.
\newblock {\em Trends in Ecology and Evolution 11}, 4 (1996), 158--163.

\bibitem{Webbetal_2009}
{\sc Webb, A., Hancock, J.~M., and Holmes, C.~C.}
\newblock {Phylogenetic inference under recombination using Bayesian stochastic
  topology selection}.
\newblock {\em {Bioinformatics} {25}}, {2} ({2009}), {197--203}.

\bibitem{Xietal_2012}
{\sc Xi, Z., Ruhfel, B.~R., Schaefer, H., Amorim, A.~M., Sugumaran, M.,
  Wurdack, K.~J., Endress, P.~K., Matthews, M.~L., Stevens, P.~F., Mathews, S.,
  and Davis, C.~C.}
\newblock {Phylogenomics and a posteriori data partitioning resolve the
  Cretaceous angiosperm radiation Malpighiales}.
\newblock {\em {PNAS} 109}, 43 (2012), 17519--17524.

\bibitem{Xieetal_2011}
{\sc Xie, W., Lewis, P.~O., Fan, Y., and Chen, M.-H.}
\newblock {Improving Marginal Likelihood Estimation for Bayesian Phylogenetic
  Model Selection}.
\newblock {\em Systematic Biology 60\/} (2011), 150--160.

\bibitem{Yang1993}
{\sc Yang, Z.}
\newblock Maximum-likelihood estimation of phylogeny from {DNA} sequences when
  substitution rates differ over sites.
\newblock {\em Molecular Biology and Evolution 10}, 6 (1993), 1396--1401.

\bibitem{Yang1994a}
{\sc Yang, Z.}
\newblock Maximum likelihood phylogenetic estimation from {DNA} sequences with
  variable rates over sites: approximate methods.
\newblock {\em Journal of Molecular Evolution 39\/} (1994), 306--314.

\bibitem{Yang1995}
{\sc Yang, Z.}
\newblock A space-time process model for the evolution of {DNA} sequences.
\newblock {\em Genetics 139\/} (1995), 993--1005.

\bibitem{Yang2006}
{\sc Yang, Z.}
\newblock {\em Computational {Molecular Evolution}}.
\newblock Oxford Series in Ecology and Evolution. Oxford University Press,
  Great Britain, 2006.

\bibitem{Yang2007}
{\sc Yang, Z.}
\newblock {PAML 4: Phylogenetic Analysis by Maximum Likelihood}.
\newblock {\em Molecular Biology and Evolution 24\/} (2007), 1586--1591.

\end{thebibliography}

\end{document}